# The origin of grain size effects in Ba(Ti$_{0.96}$Sn$_{0.04}$)O$_3$ perovskite ceramics with superior electrical properties

Yongqiang Tan,[1,2] Giuseppe Viola,[3,4] Vladimir Koval,[5] Chuying Yu,[3] Amit Mahajan,[3] Jialiang Zhang,[2] Haibin Zhang,[1,*] Xiaosong Zhou,[1] Nadezda V. Tarakina,[3] and Haixue Yan[3,*]

[1]*Innovation Research Team for Advanced Ceramics, Institute of Nuclear Physics and Chemistry, China Academy of Engineering Physics, Mianyang, 621900, China*

[2]*School of Physics, State Key Laboratory of Crystal Materials, Shandong University, Jinan, 250100, China*

[3]*School of Engineering and Materials Science, Queen Mary University of London, Mile End Road, London, E1 4NS, United Kingdom*

[4]*Department of Applied Science and Technology, Institute of Materials Physics and Engineering, Corso Duca Degli Abruzzi 24, 10129 Torino, Italy*

[5]*Institute of Materials Research, Slovak Academy of Sciences, Watsonova, 47, 040 01 Kosice, Slovakia*

The study of grain size effects in ferroelectric ceramics has attracted great research interest over the last 50 years. Although different theoretical models have been proposed to account for the variation in structure and properties with grain size, the underlying mechanisms are still under debate, creating a significant level of uncertainty in the field. Here, we report the results of a study on the influence of grain size on the structural and physical properties of Ba(Ti$_{0.96}$Sn$_{0.04}$)O$_3$, which represents a model perovskite system, where the effects of point defects, stoichiometry imbalance and phase transitions are minimized by Sn substitution. It was found that different microscopic mechanisms are responsible for the various grain size dependences observed. In fine-grained ceramics, high permittivity is due to high domain wall density and polar nanoregions; high d$_{33}$ in coarse-grained ceramics results from a high degree of domain alignment during poling; large electric field-induced strain in intermediate-grained ceramics is an outcome of a favourable interplay between constraints from grain boundaries and reversible reorientation of non-180º domains and polar nanoregions. These paradigms can be regarded as general guidelines for the optimization of specific properties through grain size control.

## I. INTRODUCTION

The study of grain size effects in polycrystalline ferroelectrics has drawn great interest in the past decades and it still represents a hot topic in the area of ferroelectrics [1-11]. The majority of studies on grain size effects currently available in the literature are mainly focused on two aspects. One concerns the investigation of ferroelectric properties in nanometer-scale structures and it is devoted to clarifying the size limits of ferroelectricity [12-16]. The other aspect is mainly related to the understanding of the grain size dependence of dielectric, piezoelectric and ferroelectric/ferroelastic behavior in bulk systems, aimed at obtaining desired properties by appropriate tailoring the grain size [4,6,9,10,17-25].

Regarding the grain size effects on dielectric properties, barium titanate (BaTiO$_3$) is probably the most widely studied material among ferroelectric ceramics [2,4-6,8,17,18, 23,24,26,27]. The permittivity of BaTiO$_3$ ceramics generally shows a peak value around the grain size of 1μm, regardless of the raw materials and sintering methods used [8,17-18]. The origin of the permittivity peak has been commonly attributed to the maximum density

______________

**Corresponding Authors**:

*Haibin Zhang. Email: hbzhang@caep.cn
*Haixue Yan. Email: h.x.yan@qmul.ac.uk

and mobility of 90º domain walls [6,8,17-18,24,27].

Although grain size effects in ferroelectrics have been extensively studied, a satisfactory understanding of all the trends observed has not been achieved yet. In fact, large discrepancies in the grain size dependences of piezoelectric properties have been reported for different ferroelectric systems [6,8-10,19,21-25,27]. Even for pure $BaTiO_3$ ceramics, just the use of different raw materials and different sintering methods may lead to completely different grain size dependencies of the piezoelectric coefficients [6,8,23,24,26,27]. The origin of the diverse grain size dependencies can be ascribed to intrinsic factors, such as crystal structure and phase coexistence, as well as extrinsic factors, including the degree of domains' alignment and the presence of charged point defects [8,21,28,29]. In particular, a significant influence of point defects on grain size effects has been recently demonstrated by comparing the grain size effects in $BaTiO_3$ ceramics prepared by pressureless sintering and spark plasma sintering (SPS) [8].

In order to fully understand the effects of grain size in ferroelectrics, high purity systems, with a high degree of stoichiometric balance and with the minimal presence of defects, are needed. During our compositional screening, it was noticed that the formation of point defects in $BaTiO_3$ ceramics during sintering at high temperatures can be suppressed by the substitution of $Ti^{4+}$ with $Sn^{4+}$ or $Zr^{4+}$ ions [30,31]. Besides that, the addition of the isovalent Sn or Zr ions allows for shifting the temperatures of the phase transitions from the rhombohedral *R3m* to the orthorhombic *Amm2* phase and from the orthorhombic *Amm2* to the tetragonal *P4mm* structure [32-37]. This enables studying grain size effects in stable phases, sufficiently far from phase transition temperatures.

In this paper, $Ba(Ti_{0.96}Sn_{0.04})O_3$ (abbreviated as BTS hereafter) ceramics were prepared to study the effect of grain size on the phase, crystal structure, domain structure, dielectric permittivity, piezoelectric $d_{33}$ coefficient and ferroelectric/ferroelastic properties with minimal influence of point defects, stoichiometry imbalance and phase transition critical behavior. The results obtained contribute to a more comprehensive understanding of the grain size effects in ferroelectric/ferroelastic bulk ceramics, knowledge that enable preparing the ceramics with improved dielectric, piezoelectric and ferroelectric/ferroelastic properties.

## II. EXPERIMENTAL

### A. Sample preparation

$Ba(Ti_{0.96}Sn_{0.04})O_3$ ceramics were prepared from $BaCO_3$ (SINOPHARM, China, purity ≥ 99.0%), $TiO_2$ (SINOPHARM, China, purity ≥ 99.8%) and $SnO_2$ (SINOPHARM, China, purity≥99.8%) powders, which were weighed according to the stoichiometric formula and thoroughly mixed. The homogenized mixture was then ball-milled for 12 hrs in alcohol and calcined at 1050°C for 4 hrs. After calcination, the BTS powder was ball-milled again for 12 hrs. Dense BTS ceramics with different grain sizes were sintered using a commercial SPS furnace (HPD-25/1 FCT Systeme GmbH) at different temperatures in the range 1160-1320°C (1160°C, 1200°C, 1240°C, 1280°C and 1320°C) under a uniaxial pressure of 85 MPa for 5 minutes. The ceramics are labeled as BTS1160, BTS1200, BTS1240, BTS1280, and BTS1320 throughout the manuscript. The SPS-processed samples (denoted as SPSed hereafter) were further annealed in air at 1000°C for 2 hrs in order to remove the carbon diffused in the samples during SPS sintering.

### B. Microstructural characterizations

The microstructure of the ceramics was studied using scanning and transmission electron microscopy. Scanning electron microscopy (SEM) was performed on polished and chemically etched samples using a JEOL JSM 6300 scanning electron microscope. The etching was performed in an aqueous solution of 5% HCl with a small amount of HF (3 drops of HF: 20 ml HCl solution) for 10 seconds. Transmission electron microscopy (TEM) was performed using a JEOL JEM 2010 microscope operated at 200 kV. Thin lamellas for TEM observations were prepared by focused ion beam milling using an FEI

Quanta 3D Dual Beam system (operational conditions: 30 kV Ga+ ion beam and currents down to 28 pA). When the thickness of the lamellas was about 100 nm, a final cleaning of the specimens was performed at 5 kV and 16 pA.

The local out-of-plane piezoresponse was measured using a scanning probe microscope (Ntegra Prima, NT-MDT) equipped with an external lock-in amplifier (SR 830, Stanford Research). A conductive probe (MikroMasch HQ: NSC15/Pt) working at the resonance frequency of 325 kHz and a force constant of 40 N/m was used for the acquisition of the piezoresponse data. The signals were recorded at 10 V and 50 kHz.

### C. XRD tests and refinements

The crystal structure of the samples was studied by X-ray diffraction (XRD). XRD data were collected on an automated Rigaku X-ray diffractometer (model Ultima IV) with Cu $K_\alpha$-radiation in the 2θ range 20° ≤ 2θ ≤ 60° and at temperatures -100°C, 25°C, 75°C and 147°C. The Rietveld refinement of the crystal structure was performed with the FullProf software package [38,39]. The experimental profiles were fitted with the Thompson-Cox-Hastings pseudo-Voigt analytical function. The goodness of fit was used as numerical criteria to assess the quality of the diffraction data fitting.

### D. Electrical characterizations

For the electrical characterization, ceramic specimens were coated on the top and bottom surfaces with silver paste, which was fired at 600°C for 20 minutes. The relative dielectric permittivity was measured using a precision impedance analyser (Agilent, a model 4294A) connected to a chamber with controlled temperature. The room-temperature polarization-electric field (P-E) hysteresis loops and strain-electric field (S-E) loops were measured with triangular electric field waveforms of 3 kV/mm amplitude and 1 Hz frequency using a ferroelectric tester (NPL, Teddington, UK) [40]. The poling was carried at 95°C in silicon oil under a dc field of 3 kV/mm for 30 min. The piezoelectric $d_{33}$ coefficient was measured using a Berlincourt-type $d_{33}$ meter (YE 2730A, China).

## III. RESULTS AND DISCUSSIONS

Fig. S1a - e show the micrographs of the polished and chemically etched surfaces of the BTS ceramics sintered at five different temperatures. All the samples investigated have low porosity, and consistently high relative densities (> 98%) as measured by the Archimedes method. The average grain size increases from 0.8μm to 12.5μm as the sintering temperature increases from 1160°C to 1320°C (Fig. S1f). In fine-grained BTS ceramics, the typical stripe-like domain structure, commonly observed in fine-grained $BaTiO_3$ ceramics [8], is not clearly recognized (Fig. S1a and S1b). It appears that the grains contain separated domain islands instead of stripe domains. The peculiar domain morphology may result from a disruption of stripe-type domains by the reducing grain size. As the grain size increases, the presence of domains becomes increasingly neater (Fig. S1c-S1e). In coarse-grained BTS ceramics, the domain patterns are clearly visible (Fig. S1d and S1e). The watermark-type regions correspond to the 180° domains, while the stripe-type areas represent non-180° domains [41,42].

The presence of domains in micrometer-sized grains of the BTS1160 sample is evidenced by TEM (Fig. 1 and S2). Fig. 1a shows the dark-field TEM image of an area of several grains. From the image contrast, one can see that the central grain consists of three domains: two twin domains (areas marked by "A" and "D") and a domain between them (corresponding to the bright area between "B" and "C"). The latter domain shows superstructure reflections in the selected-area diffraction (SAED) pattern. The superstructure can be described in a rhombohedral lattice with the unit cell parameters (in hexagonal settings) $a$ = 6.5 Å and $c$ =12.7 Å (Fig. 1B and 1C). The two twin domains are rotated by 30° with respect to each other, as evidenced in the corresponding SAED patterns (Fig. 1A-D). A similar domain structure is observable also on the right side of the grain located in the bottom left corner of Fig.1a and 1b (the area marked by "F" in Fig.1b and 1F). However, the SAED pattern from the left side of the same grain (marked by "E" in Fig.1b and 1E) presents extra spots along the [1-10] direction of the crystal. In both cases, the formation of different superstructures is most probably related to local cation ordering controlled by an inhomogeneous distribution of Sn [32]. Since the difference in the unit cell parameters of the orthorhombic BTS (space group *Amm*2) and the tetragonal BTS (space group *P4mm*) unit cells is less than 0.01 Å, the structure cannot be unambiguously distinguished from the SAED patterns (the indexing of the SAED patterns has been performed based on the tetragonal unit cell).

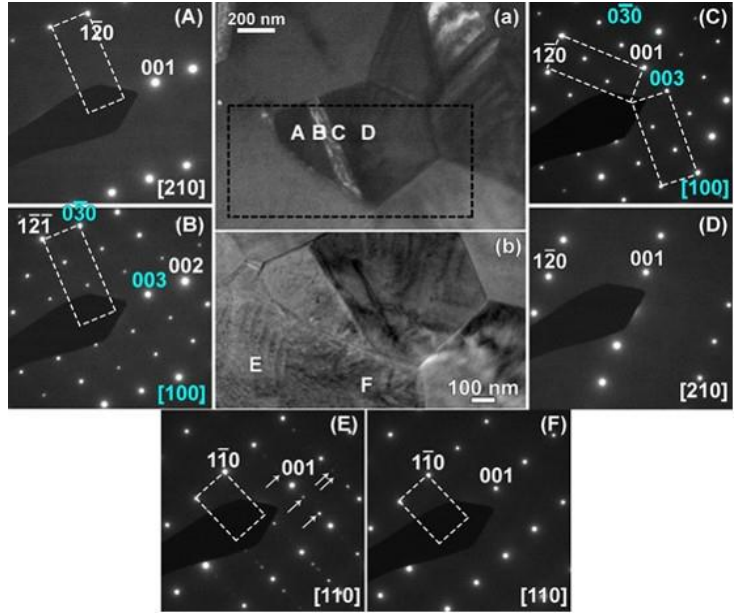

FIG. 1. (a) Dark-field TEM image of BTS ceramics sintered at 1160°C. A, B, C and D denote the positions at which the SAED patterns shown in (A), (B), (C) and (D), respectively, were obtained. (b) Bright-field transmission electron microscopy image of Sn-doped BaTiO$_3$ obtained from the area marked by the black dotted line in (a). E and F denote positions at which the SAED patterns shown in (E) and (F), respectively, were obtained. All SAED patterns were indexed in the tetragonal unit cell (white indices), patterns (B) and (C) were additionally indexed in a rhombohedral unit cell (blue indices). Extra spots on SAED pattern (E) are marked with white arrows.

Piezoresponse force microscopy (PFM) was used to investigate the domain structure in BTS1160 and BTS1280. The dark and bright areas in the phase images (Fig. 2) correspond to domains with polarization oriented parallel (marked by green circles in Fig. 2b) and antiparallel (marked by blue circles in Fig. 2b) to the applied field, respectively. BTS1160 shows a mosaic-like pattern with sub-micrometer sized domains along with the possible presence of polar nanoregions (PNRs) as illustrated by the phase image in Fig. 2c and the corresponding line profile in Fig. 2d. These images provide further details of the domain patterns observed in the micrographs (Fig.S1a and S1b). The mosaic-like domain configuration and PNRs resembles the patterns characteristic of relaxor ferroelectrics [43]. The observation of a mosaic-like pattern clearly proves that even the BTS ceramic with the smallest grains here studied presents a multi-domain configuration within a single grain. The phase images of BTS1280 (Fig. 2f and 2g) demonstrate the presence of stripe-like domains, which is in accordance with the SEM observations (Fig. S1d). The average domain size, as determined from the line profile in Fig. 2h (corresponding topography is presented in Fig. 2e), is ~ 1.4μm.

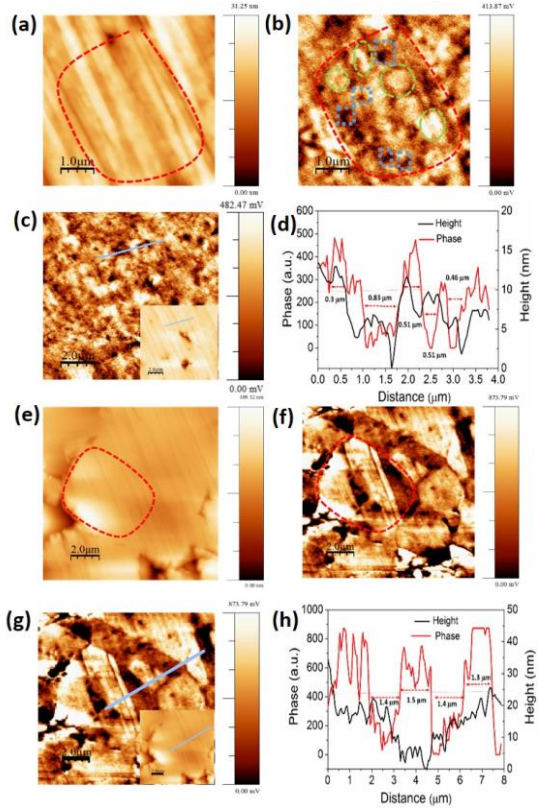

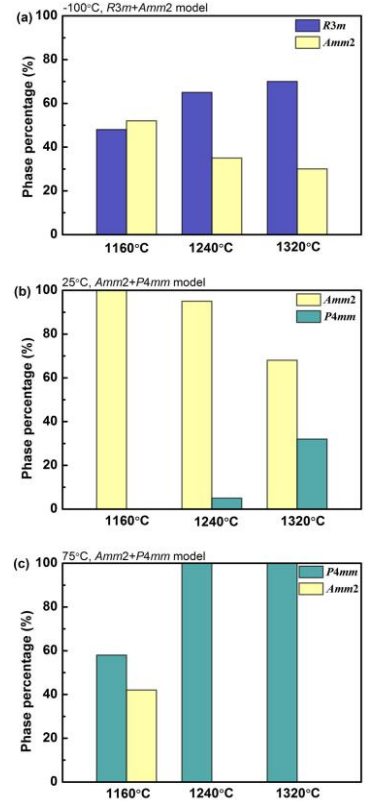

FIG. 2. (a) Topography of polished surface of BTS1160 ceramic; (b) PFM phase images of BTS1160 ceramic; (c) Selected area for the PFM line scanning and (d) corresponding line profile analyses for the BTS1160 ceramic; (e) Topography of polished surface of the BTS1280 sample; (f) PFM phase images of BTS1280; (g) Selected area for the PFM line scanning and (h) corresponding line profile analyses for the BTS1280 ceramic.

Fig. 3 presents the phase compositions of the BTS ceramics with small (BTS1160), intermediate (BTS1240) and large (BTS1320) grains. The refined lattice parameters and $R$-factors, as obtained from the Rietveld analysis of the XRD data measured at four different temperatures (Fig. S3), are summarized in Table S1 (in supplementary material). At -100°C, all the samples (BTS1160, BTS 1240 and BTS 1320) present the coexistence of rhombohedral ($R3m$) and orthorhombic ($Amm2$) phases (Fig. 3a), with the latter being dominant in the fine-grained ceramic BTS1160. At 25°C, the latter shows a pure orthorhombic structure (Fig. 3b). As the grain size increases, the amount of tetragonal phase increases, so that the coarse-grained BTS1320 ceramic appears as a mixture of the orthorhombic structure ($Amm2$) and the tetragonal phase ($P4mm$) (Fig. 3b), in accordance with previous studies on pure $BaTiO_3$ [2,35] and Sn-modified $BaTiO_3$ ceramics [34]. It can be inferred that the orthorhombic structure is favoured over the tetragonal phase in fine-grained BTS ceramics probably because it offers a higher number of directions for the spontaneous polarization, which helps to minimize the energy of the system [2]. This is also consistent with the shift of $T_{O-T}$ towards higher temperatures with decreasing grain size in $BaTiO_3$ ceramics [2, 8]. At 75°C, BTS1160 exhibits the coexistence of orthorhombic and tetragonal phases; however, with increasing grain size, the orthorhombic phase disappears in favour of the tetragonal symmetry (Fig. 3c). The Rietveld analysis of the XRD data collected at 147°C suggests the presence of a tetragonal/pseudocubic phase alongside the main cubic structure ($Pm$-$3m$ space group symmetry) in all samples investigated. Refinements with a bi-phasic ($Pm$-$3m$+$P4mm$) model gave better values of the $R$-factors than those with the single cubic model (e.g., $R_p/R_{wp}$ is 7.5/10.7 and 8.2/11.1 for the $Pm$-$3m$+$P4mm$ structural model and the $Pm$-$3m$ model, respectively, for the BTS1160 sample). In addition, the goodness of fit (GOF) factor was substantially improved when the mixed structural model was employed (for BTS1160; GOF ($Pm$-$3m$+$P4mm$) ~ 1.5 vs. GOF ($Pm$-$3m$) ~ 2.1).

FIG. 3. Phase composition of the BTS ceramics at (a) -100°C, (b) 25°C and (c) 75°C, as obtained from the Rietveld refinement of the XRD data.

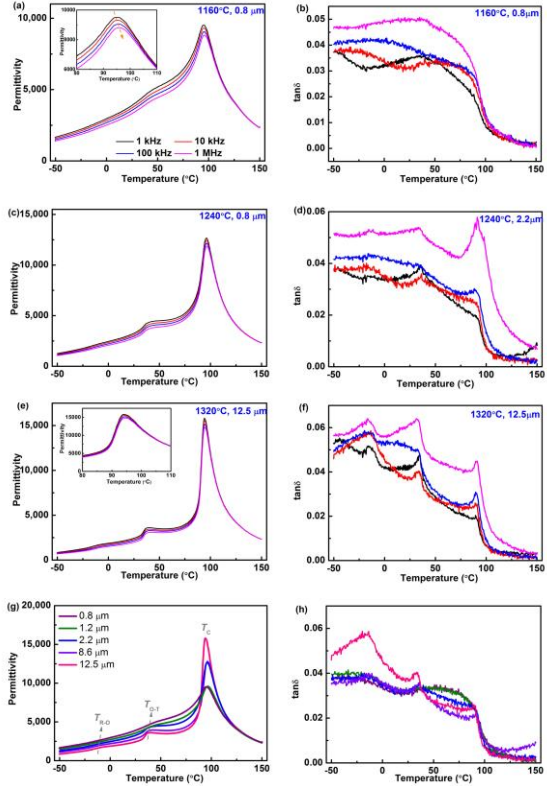

FIG. 4. (a-f) The temperature dependence of the relative dielectric permittivity and loss of BTS ceramics sintered at three different temperatures (a, b - 1160°C; c, d - 1280°C and e,f - 1320°C) and measured at 1 kHz, 10 kHz, 100 kHz and 1 MHz. (g, h) The temperature dependence of the relative dielectric permittivity and loss of BTS ceramics with different grain size (measured at 1kHz).

Fig. 4a - f show the temperature dependences of the relative dielectric permittivity and loss of unpoled BTS ceramics with different grain size. Similarly to pure BaTiO$_3$ ceramics, three dielectric anomalies can be identified in the temperature range investigated (-50°C-150°C), which correspond to the rhombohedral-to-orthorhombic ($T_{R-O}$),

orthorhombic-to-tetragonal ($T_{O-T}$) and tetragonal-to-cubic ($T_C$) phase transitions. The dielectric anomalies associated with the phase transitions are much more pronounced in coarse-grained samples (Fig. 4e and 4f), while for fine-grained ceramics (Fig. 4a and 4b) only the Curie point $T_C$ can be clearly identified. From the temperature dependence of the relative dielectric permittivity, it was difficult to determine the grain size dependence of $T_{R-O}$. More obvious sensitivity to a variation in grain size can be observed in the structural transformation temperature from the orthorhombic to the tetragonal phase. $T_{O-T}$ gradually shifts to a higher temperature by decreasing the average grain size. The Curie point $T_C$ does not show a significant grain size dependence when the grains are larger than 2.2μm, which is consistent with the observation that the tetragonality is nearly grain size independent (**Table S1**). In comparison with pure BaTiO$_3$ [8], all the BTS samples show higher $T_{R-O}$ and $T_{O-T}$ temperatures, while lower $T_C$.

In BTS1160, the permittivity peak around $T_C$ broadens and it slightly shifts towards higher temperatures with increasing frequency (see inset of Fig. 4a). These observations are in agreement with previous data on Ba(Ti$_{1-x}$Sn$_x$)O$_3$ ceramics with x > 0.19 [44, 45]. In BTS1320, the permittivity peak at $T_C$ shows a negligible dependence on the frequency (inset of Fig. 4e). This suggests that by decreasing the grain size the paraelectric-ferroelectric phase transition becomes diffused and that the dielectric behaviour begins to show relaxor-like characteristics, as previously reported for fine-grained BaTiO$_3$ ceramics [46]. The relaxor-like behaviour in fine-grained BTS ceramics can be attributed to the presence of PNRs, whose formation could be due to an intrinsic disturbance of translational symmetry, consistently with the inhomogeneous distribution of Sn$^{4+}$ observed in our TEM study, also reported in previous investigations [32]. Shi et al. [32] have demonstrated that PNRs can destabilize the long-range ordering and suppress the formation of typical domain structures in ferroelectrics. Due to the slightly larger radius of the Sn ions (69 pm) compared to that of the Ti ions (61 pm) [32], $Sn^{4+}$ has less available space to shift within the oxygen octahedron; therefore, it experiences a smaller off-centre displacement [47]. In fine-grained BTS ceramics, the higher internal stress from grain boundaries further suppresses the off-centre displacement of $Sn^{4+}$, giving rise to PNRs which are probably responsible for the relaxor-like characteristics observed. On the other hand, in BTS ceramics with coarse grains, there is less constraint to the off-centre displacement of $Sn^{4+}$ and $Ti^{4+}$, thus the presence of macroscopic domains prevails over that of PNRs.

When the temperature is above $T_C$, the permittivity of BTS ceramics shows no grain size dependence (Fig. 4g). At $T_C$, where the contribution of domain walls density to the net polarization usually vanishes, the dielectric permittivity decreases with decreasing grain size (Fig. 4g), as a result of the increased number of grain boundaries with low permittivity. Upon cooling through the paraelectric-ferroelectric phase transition, the dielectric permittivity shows clear grain size dependence. In all ferroelectric phases, the permittivity increases with decreasing the average grain size down to 0.8μm (Fig. 4g). This can be attributed to the increased presence of domain walls and PNRs in the ceramics with small grain size.

The dielectric loss above $T_C$ of all BTS ceramics was found to be less than 0.01, suggesting that the formation of point defects during sintering at high temperatures was hindered by $Sn^{4+}$ doping. Therefore, the influence of point defects on the grain size dependences observed in this study can be neglected. The reduced concentration of point defects in BTS ceramics is also demonstrated by the temperature dependence of the permittivity measured up to 800°C (**Fig. S4**). The permittivity peak around 600°C, often related to the presence of oxygen vacancies in perovskites [48], was significantly suppressed even in the BTS ceramic sintered at 1320°C, if compared to the dielectric behaviour of pure $BaTiO_3$ ceramics [8]. It has been previously reported that the presence of oxygen vacancies in $BaTiO_3$ is strongly related to the unstable valence state of $Ti^{4+}$ [49]. A partial substitution of $Ti^{4+}$ with more chemically-stable $Zr^{4+}$ was theoretically and experimentally proved to decrease the amount of oxygen vacancies in $BaTiO_3$ [30,31]. Taking in account the similarities between the chemical stability of $Zr^{4+}$ and $Sn^{4+}$ cations [34], the inhibiting effect of $Sn^{4+}$ addition on the formation of oxygen vacancies in BTS can be explained.

The polarization-electric field (P-E) loops and the current-electric field (I-E) curves of BTS ceramics sintered at three different temperatures (1160°C, 1240°C and 1320°C) are shown in Fig. 5. Fig. 5a, 5b and 5c show the P-E and I-E loops traced at room temperature, while those recorded at 125°C are presented in Fig. 5d, 5e and 5f. It can be seen that at room temperature, all the BTS samples show typical ferroelectric P-E loops. With increasing grain size, the loops become increasingly more saturated. The values of $P_{max}$ and $P_r$ of BTS1160 (average grain size 0.8μm) are 20.4 μC/cm$^2$ and 7.9 μC/cm$^2$, respectively. As the grain size increases from 0.8μm to 12.5μm, $P_{max}$ slightly increases from 20.4 μC/cm$^2$ to 21.3 μC/cm$^2$, and $P_r$ sharply increases to 14.5 μC/cm$^2$, which is almost twice the value found in BTS1160 ceramics (7.9 μC/cm$^2$). The observed dependences of $P_{max}$ and $P_r$ on the grain size are in good agreement with those earlier reported for the $BaTiO_3$ ceramics prepared by SPS from nano-sized powders [8, 50]. The coercive field continuously decreases with increasing grain size from 0.314 kV/mm in BTS1160 to 0.138 kV/mm in BTS1320.

It is worth mentioning that all ceramics show signs of domain switching even at 125°C (> $T_C$ ~ 105°C), as evidenced by the tiny current peaks in the I-E curve (Fig. 5d-5f), which can be attributed to the reorientation of PNRs above $T_C$ (the *P4mm* phase in Table S1) [51, 52]. Eventually, the first-order ferroelectric/paraelectric phase transition characteristic of $BaTiO_3$-based systems could also be partially responsible for the switching-like event detected above $T_C$ [51]. In coarse-grained BTS1320 ceramics, the domain switching current peak above $T_C$ becomes clearly suppressed (Fig. 5f). The temperature dependences of the maximum and remanent polarization of BTS1160, BTS1240 and BTS1320 are shown in Fig. 5g. With increasing temperature, $P_{max}$ and $P_r$ slightly decrease until 50°C; at higher temperature, a significant decrease of both quantities is observed in all the three ceramics investigated. It is interesting to note that the dependence of both $P_{max}$ and $P_r$ on the grain size varies within the temperature range considered. The maximum polarization $P_{max}$ slightly decreases with decreasing grain size until 50°C-75°C; however, at higher temperatures, fine-grained ceramics start showing larger $P_{max}$. Analogous observations can be made for $P_r$, which markedly decreases with decreasing grain size until 75°C; at T>75°C, $P_r$ starts to increase with decreasing grain size (Fig. 5g). Above $T_C$, the value of $P_{max}$ of BTS1160 is higher than that of BTS1280 and BTS1320. This further confirms the presence of PNRs and their dominant contribution to the polarization in fine-grained ceramics at $T > T_C$.

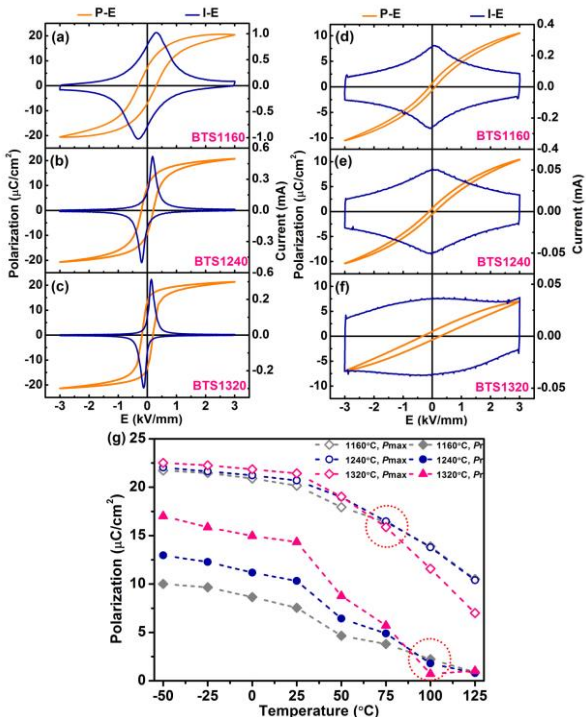

FIG. 5. Polarization-electric field and current-electric field curves for BTS1160, BTS1240 and BTS1320 measured at (a-c) 25°C and (d-f) 125°C; Temperature dependences of $P_{max}$ and $P_r$ for BTS ceramics sintered at different temperatures (g).

Fig. 6a presents the room-temperature strain-electric field (S-E) loops (without including the first poling cycle) of BTS ceramics with different grain size. It can be seen that all the S-E curves exhibit a typical butterfly-like shape with symmetrical wings, even for the samples sintered at high temperature. This is quite different from the bipolar S-E curves shown by pure $BaTiO_3$ ceramics, which evidenced a remarkable tendency to become asymmetrical when the sintering temperature is higher than 1200°C [8, 53]. The asymmetry is commonly attributed to the development of an internal bias due to hindering of the domain wall movement by point defects [8, 53-55]. The symmetrical S-E loops in BTS ceramics may further validate the hypothesis that a small amount of $Sn^{4+}$ addition might suppress the formation of point defects, such as oxygen vacancies, during sintering at high temperatures. From Fig. 6a it can be seen that with increasing grain size the strain amplitude in the bipolar test gradually increases until a maximum value of about 0.2%, observed at 2.2µm average grain size, beyond which the strain amplitude drastically decreases. Fig. 6b shows the unipolar S-E curves of dc-poled BTS ceramics with different grain size. The measurements were carried out by setting as zero the remanent strain after the poling process. All dc-poled BTS ceramics show slim unipolar S-E loops with a nearly zero remanent strain. Similarly to the bipolar strain curves, the maximum strain achieved in the unipolar test first increases and then decreases with increasing grain size, showing a peak value of 0.23% at 2.2µm grain size. The strain-polarization (S-P) loops of BTS ceramics with three different grain size (0.8µm, 2.2µm and 12.5µm) are shown in Fig. 6c. BTS1160 (0.8µm grain size) shows a slim S-P loop without hysteresis, indicating that in the fine-grained BTS ceramic the dominant contribution to both polarization and strain is given by non-180º domains or PNRs [40,56]. The S-P loop of BTS1280 with 2.2µm grains shows a small hysteresis, suggesting that non-180º domains or PNRs still play a dominant role [40]. For the coarse-grained BTS1320 ceramic with 12.5μm grain size, the hysteresis of the S-P loop becomes prominent. Additionally, the polarization becomes larger and the strain significantly reduces, indicating that the extent of 180º domain switching is much larger than that occurring in fine-grained BTS ceramics. Fig. 7 summarizes the grain size dependencies of the dielectric and piezoelectric characteristics. The dependence of the relative dielectric permittivity on grain size at three specific temperatures (-50°C, 25°C and 50°C) in Fig. 7a was extrapolated from Fig. 4. The permittivity exhibits similar grain size dependencies in the three different phases: it keeps increasing with decreasing grain size without showing any peak in the grain size range studied. This is different from $BaTiO_3$ ceramics, which show a permittivity peak around 1μm grain size [6, 8, 17, 24, 57]. The room-temperature dielectric permittivity of BTS1320 with an average grain size of 12.5μm is about 2400, while it increases to about 3900 when the average grain size is decreased to 0.8μm. The permittivity of all BTS ceramics decreases after dc poling; the decrease is larger in the coarse-grained sample (see Fig. 7b). This can be explained by the lower domain wall density in coarse-grained ceramics after poling [8].

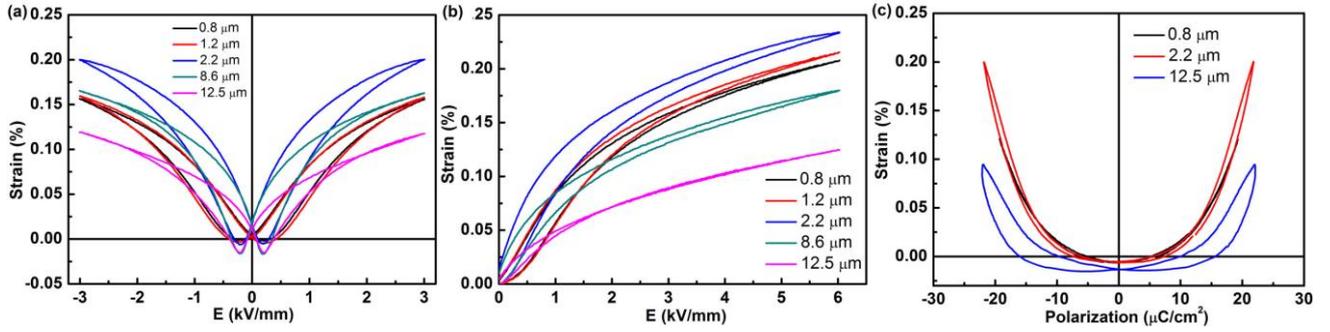

FIG. 6. Bipolar (a) and unipolar (b) electric field-induced strain behaviour of BTS ceramics with different grain sizes; (c) strain vs. polarization curves for BTS ceramics with different grain sizes.

The grain size dependence of the piezoelectric coefficient $d_{33}$ is shown in Fig. 7c. One can see that it exhibits a totally opposite grain size dependence compared to that of the relative dielectric permittivity at all the three temperatures considered. In ceramics with an average grain size of 0.8μm, the $d_{33}$ at room temperature is 100 pC/N, and it gradually increases with increasing the average grain size. When the average grain size is 12.5μm, the room-temperature $d_{33}$ reaches 380 pC/N, which is almost four times larger than that of the fine-grained BTS ceramic. The grain size dependences of the $d_{33}$ at different temperatures show similar trends. It should be noted that the piezoelectric coefficient increases with increasing temperature, approaching the orthorhombic-tetragonal transition ($T_{O-T}$ ~ 40°C). Thermal depoling effects become significant only above 75°C.

The room-temperature piezoelectric coefficient $d_{33}^*$ of poled BTS ceramics was calculated as the ratio $S_{max}/E_{max}$ [58] using the data in Fig. 6b. The grain size dependence of $d_{33}^*$ is presented in Fig. 7d. One can see that the $d_{33}^*$ first increases with increasing the average grain size, then it reaches a maximum (~ 390 pm/V) around 2μm, and it finally decreases with a further increase of the grain size, until a value of about 200 pm/V in BTS ceramics with an average grain size of 12.5μm. The grain size dependence of the electric field-induced strain of BTS ceramics is quite similar to that of pure $BaTiO_3$ ceramics prepared from 100 nm $BaTiO_3$ powders by SPS [59].

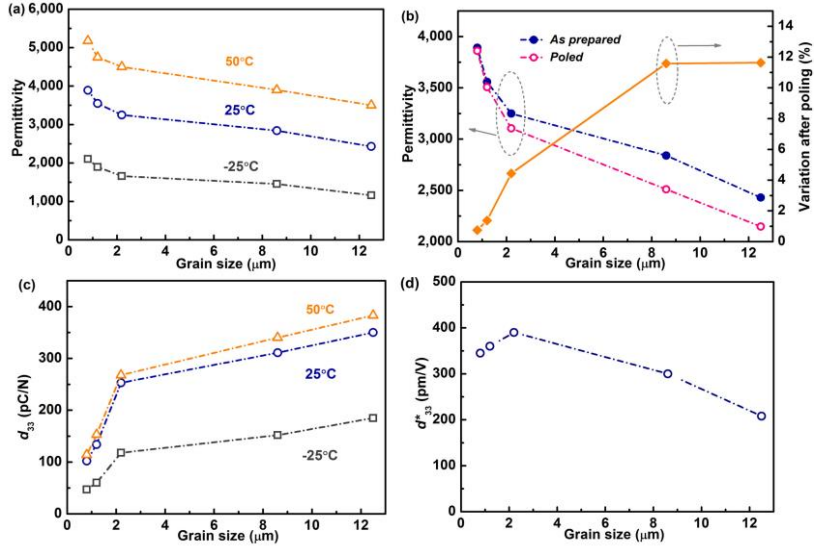

FIG. 7. (a) The grain size dependence of the permittivity of as-prepared (unpoled) BTS ceramics at different temperatures; (b) the grain size dependence of the permittivity of as-prepared (unpoled) and poled BTS ceramics at room temperature; (c) the grain size dependence of $d_{33}$ of BTS ceramics at different temperatures; (d) the grain size dependence of $d_{33}^*$ of as-prepared (unpoled) and poled BTS ceramics at room temperature.

From the obtained experimental results, it can be summarised that in BTS ceramics, where point defects do not have a significant influence on the trends observed, the dielectric permittivity, the piezoelectric coefficient and the electric field-induced strain show different grain size dependences. Fine-grained samples exhibit high dielectric permittivity; coarse-grained samples show higher $d_{33}$; samples with intermediate grain size can experience larger electric field-induced strain. This evidences that high dielectric permittivity, high piezoelectric constant and large electric field-induced strain in BaTiO$_3$-based ceramics originate from different physical mechanisms.

The experimental and theoretical studies of the grain size effect on the dielectric permittivity of BaTiO$_3$ ceramics have demonstrated that high permittivity is closely related to the maximized domain wall density in fine grains [6, 8, 17, 24, 46]. The observed grain size dependence of the permittivity in BTS ceramics is similar to that found in pure BaTiO$_3$ [8, 17, 24, 27], implying that the high density of domain walls plays an important role also in BTS ceramics. The main difference is that BTS ceramics show no permittivity peak in the range of grain size studied, while BaTiO$_3$ ceramics exhibit a permittivity peak around 1μm grain size [6,8,17,18,24,46]. Besides, in this study, the higher permittivity in fine-grained BTS ceramics can be also due to the presence of PNRs, which provide the main contribution to the high permittivity in relaxors [60,61]. It is possible that the presence of PNRs in BTS ceramics has lowered the critical grain size where the permittivity is maximized; this is probably why no permittivity peak was found in the grain size range investigated. Future studies shall verify this hypothesis.

The grain size dependence of piezoelectric properties is mainly influenced by the mechanisms of domain switching and domain alignment under high electric fields [8,21,35]. BTS ceramics exhibit the presence of both 180º and non-180º domains; the former are pure ferroelectric and mainly contribute to the electrical polarization, while the latter are both ferroelectric and ferroelastic, and thus they can simultaneously contribute to polarization and strain. Additionally, there is also a non-negligible contribution of PNRs to polarization and strain. From the S-P curves of BTS ceramics with different grain sizes (Fig. 6c), it can be inferred that the amount of 180º domains increases with increasing grain size within the range studied. In addition, the increased remanent polarization in coarse-grained ceramics suggests that the driving force of back switching decreases with increasing grain size. These grain size-controlled features are thought to be responsible for the increase in the $d_{33}$ coefficients of poled BTS ceramics with larger grains due to the enhanced domain alignment after the poling process.

The domain switching process under high electric field can be reversible or irreversible [62-64]. The main contribution to the electric field-induced strain in bipolar tests is given by the reversible switching of non-180º domains and possibly by PNRs, when present [56,63,65,66]. In fact, in some ferroelectric/ferroelastic ceramics, only a limited part of the strain produced during the very first electric field cycle applied on virgin samples can be recovered during field unloading and reversal [59,63]. In coarse-grained BTS ceramics, a larger portion of non-180º domain switching is irreversible. This is also evidenced in Fig.7b, which shows that the difference of the permittivity before and after poling gradually increases with increasing grain size, indicating the increased proportion of irreversible domain switching during poling. The increase of both the portion of 180º domains and the irreversible switching of non-180° domains lead to the decreased strain amplitude in coarse-grained BTS ceramics as shown in Fig. 7. In fine-grained BTS ceramics, the constraint from grain boundaries hinders domain switching [67]. Therefore, the largest amount of reversible field-induced processes occurs in ceramics with intermediate grain size, resulting from the convoluted contributions of the applied field, grain boundaries and temperature, which affect domain forward/backward switching and the stability of domains and PNRs.

## IV. Conclusion

By studying grain size effects in $Ba(Ti_{0.96}Sn_{0.04})O_3$ ceramics, the present paper identifies the grain size dependence of dielectric, piezoelectric and ferroelectric/ferroelastic properties that would be generally observed in bulk ferroelectrics, when the influence of point defects, non-stoichiometry and phase transitions is minimized. High permittivity in fine grains is achieved by maximized density of domain walls and polar nanoregions; high $d_{33}$ in coarse-grained ceramics is determined by high degree of domain alignment after poling; large electric field-induced strain in intermediate-grained ceramics is due to a favourable combination of constraints from grain boundaries and reversible reorientation of non-180º domains and eventually polar nanoregions.

## V. Acknowledgements

Yongqiang Tan would like to thank China Scholarship Council for the financial support of 1-year research in the UK. This work was financially supported by the National Natural Science Foundation of China (Grant No. 51172128 and 91326102), the Specialized Research Fund for the Doctoral Program of Higher Education (Grant No. 20130131110006) and the Science and Technology Development Foundation of China Academy of Engineering Physics (Grant No. 2013A0301012). This work was partially supported by the Grant Agency of the Slovak Academy of Sciences through a Grant no. 2/0059/17 and by the Slovak Research and Development Agency through a Project no. APVV-15-0115 (MACOMA).

**Supplementary material**

**Table S1** The refined lattice parameters of individual structural phases and reliability *R*-factors obtained from the Rietveld analysis of the XRD data as collected at -100, 25, 75 and 147ºC.

| Test temperature | Phase | Lattice parameters (Å) | | | R-factors ($R_p$, $R_{wp}$) | | |
|---|---|---|---|---|---|---|---|
| | | BTS-1160 | BTS-1240 | BTS-1320 | BTS-1160 | BTS-1240 | BTS-1320 |
| **-100°C** | *R*3m | a=5.6702(4) c=6.9441(1) | a=5.6669(0) c=6.9579(1) | a=5.6603(6) c=6.9582(1) | 6.5 | 6.5 | 10.2 |
| | *Amm*2 | a=3.9957(2) b=5.6903(5) c=5.6808(5) | a=3.9932(2) b=5.6853(6) c=5.6977(6) | a=3.9955(3) b=5.6914(7) c=5.6799(7) | 8.9 | 8.9 | 13.0 |
| **25°C** | *Amm*2 | a=3.9968(6) b=5.705(2) c=5.688(2) | a=4.00327(6) b=5.6808(2) c=5.6851(2) | a=3.9988(1) b=5.6892(4) c=5.6978(4) | 8.0 | 9.8 | 10.3 |
| | *P*4*mm* | a=4.00851(5) c=4.01965(5) | a=4.00900(4) c=4.02499(7) | a=4.00993(7) c=4.0215(1) | 10.9 | 12.9 | 13.6 |
| **75°C** | *P*4*mm* | a=4.0125(1) c=4.0239(3) | a=4.0073(1) c=4.0263(2) | a=4.0064(3) c=4.0271(1) | 7.9 | 8.0 | 8.8 |
| | *Amm*2 | a=4.0009(3) b=5.7012(1) c=5.6964(1) | | | 10.8 | 10.8 | 11.3 |
| **147°C** | *Pm-3m* | a=4.0185(1) | a=4.0179(6) | a=4.0178(5) | 7.5 | 7.1 | 8.6 |
| | *P*4*mm* | a=4.0178(3) c=4.0201(1) | a=4.0175(3) c=4.0182(2) | a=4.0179(1) c=4.0193(1) | 10.7 | 9.7 | 11.3 |

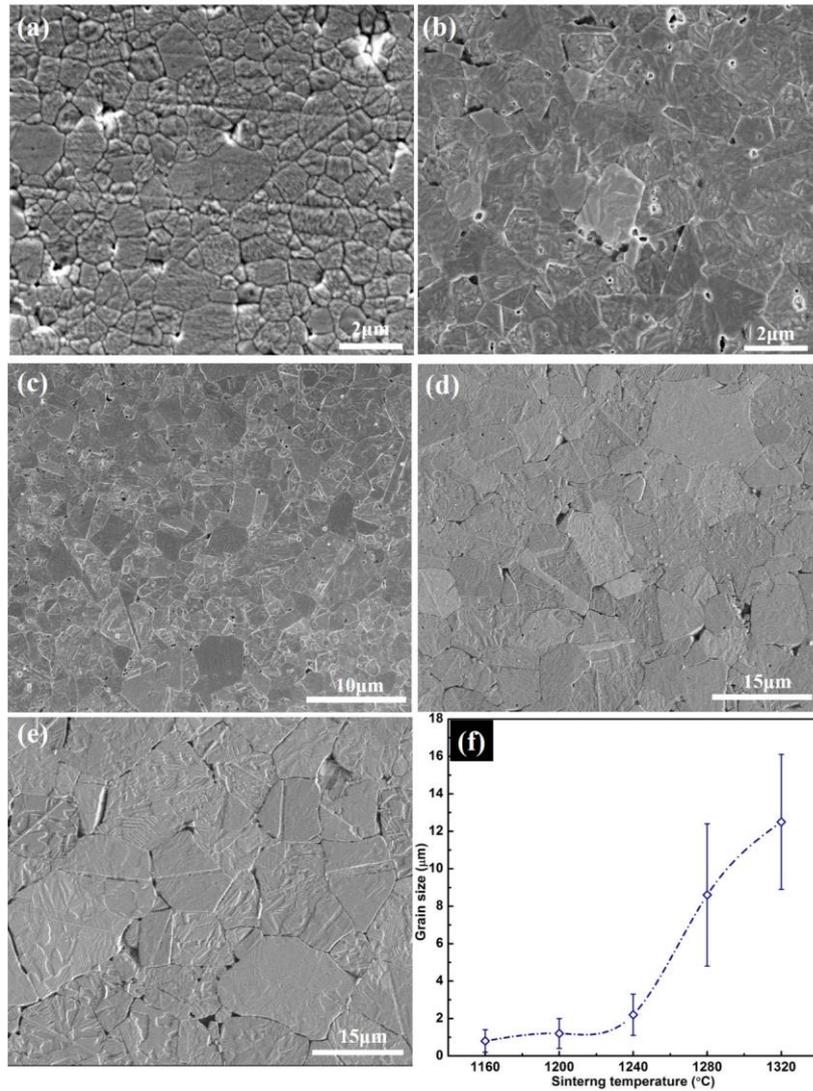

FIG. S1. SEM images of BTS ceramics sintered at (a) 1160ºC; (b) 1200ºC; (c) 1240ºC; (d) 1280ºC; and (e) 1320ºC. The average grain size, as calculated for the ceramics, is plotted as a function of the sintering temperature in (f).

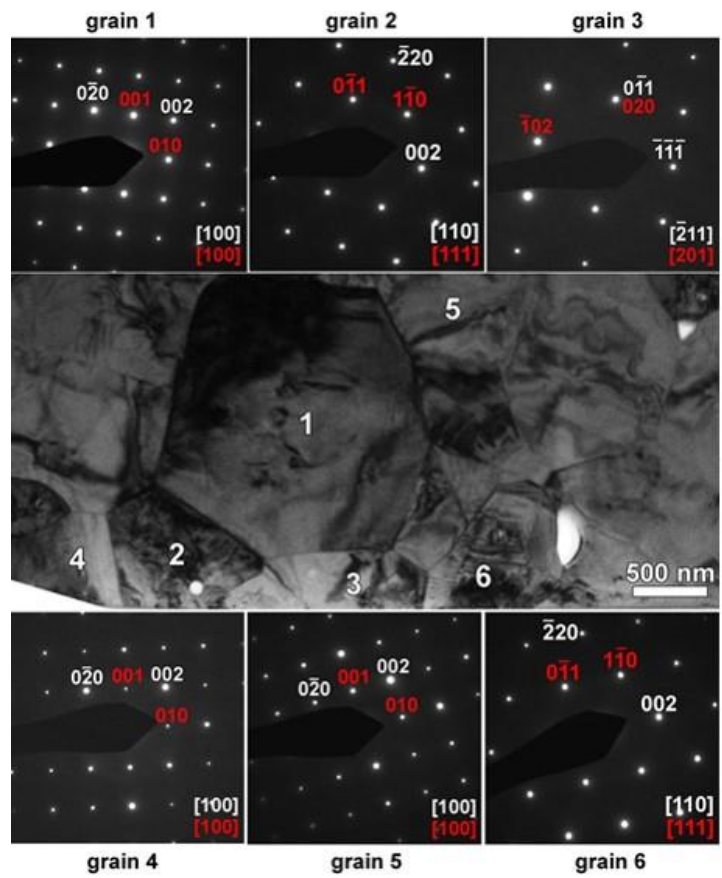

FIG. S2 Bright-field transmission electron microscopy image of BTS1160 ceramic and SAED patterns obtained from the grains marked by the numbers. All SAEDs are indexed in the orthorhombic (white indexes) and the cubic (red indexes) unit cells.

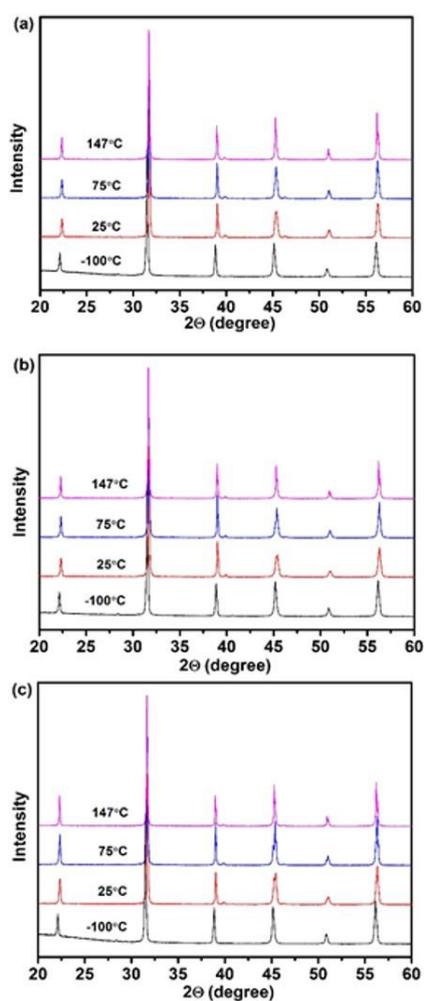

FIG. S3 Evolution of the XRD patterns with increasing temperature from -100°C to 147°C for BTS ceramics sintered at (a) 1160°C; (b) 1240°C; (c) 1320°C. (Note: Small peaks at 39.7° and 46.2° are reflections from the cubic (111) and (200) planes of the Pt holder.)

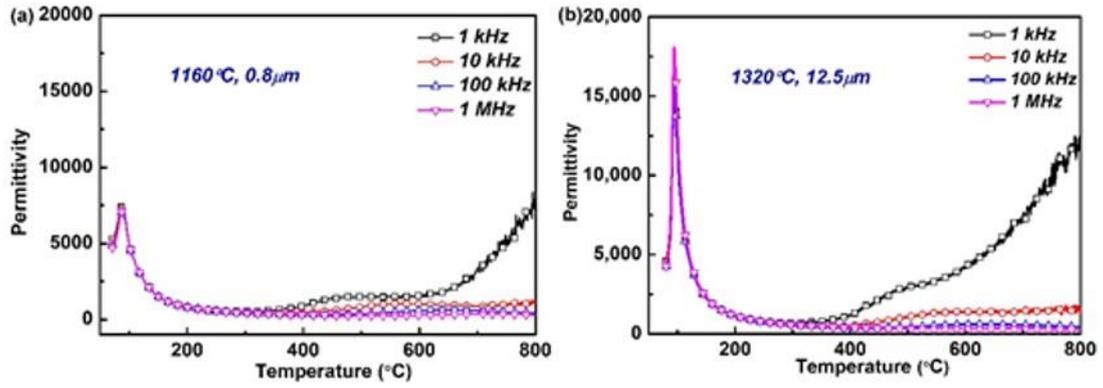

FIG. S4 The extended temperature dependence of the relative dielectric permittivity up to 800°C for BTS ceramics sintered at (a) 1160°C and (b) 1320°C. The permittivity peaks between 400°C to 600°C can be attributed to point defects. The intensity of the peaks is lower than pure $BaTiO_3$ ceramics [1,2].